\documentclass[a4paper,fleqn,usenatbib]{mnras}

\usepackage{newtxtext,newtxmath}

\usepackage[T1]{fontenc}
\usepackage{ae,aecompl}

\usepackage{graphicx}	
\usepackage{amsmath}	
\usepackage{amssymb}	

\title
[On the saturation of acoustic frequencies]
{On the saturation of acoustic mode frequencies at high solar activity}

\author[M. C. Rabello Soares]{
M. Cristina Rabello Soares, \thanks{E-mail: cristina@fisica.ufmg.br}
\\
Physics Department, Universidade Federal de Minas Gerais, Belo Horizonte, MG 31270-901, Brazil
}

\date{Accepted XXX. Received YYY; in original form ZZZ}

\pubyear{2019}

\begin{document}
\label{firstpage}
\pagerange{\pageref{firstpage}--\pageref{lastpage}}
\maketitle

\begin{abstract}
Acoustic mode frequencies 
obtained by applying spherical harmonic decomposition to HMI, MDI and GONG observations 
were analysed throughout the solar cycle.
Evidence of a deviation from a linear relation with solar radio flux
was found indicating a saturation effect at high solar activity. 
The Gompertz model,
which is one of the most frequently used sigmoid functions to fit growth data, is used.
It is shown that its fitting to MDI and GONG data are statistically significant
and a median saturation of four-hundred sfu is estimated.
This saturation level is 50\% larger than any obtained in the last century,
hence the small effect observed in the minimum-to-maximum frequency shift.
However, as shown here, it should not be disregarded.
\end{abstract}

\begin{keywords}
Sun: activity -- Sun: helioseismology -- Sun: oscillations -- stars: activity
\end{keywords}



\section{Introduction}

The Sun undergoes periods of strong and weak activity over the course of 11 years,
which is most commonly seen as a cyclic variation in the number of sunspots
present in the solar surface.
The cycles have been observed for the last 400 years
and,
starting at mid-18th century, the
cycles have been labelled with a number.
We are currently moving into cycle 25.
The origin of the eleven-year solar magnetic activity cycle is
still not completely understood.
There are yet 
questions on how the magnetic fields are built inside the Sun.
The 
results of dynamo simulations are still somewhat dependent on numerical parameters 
that are orders of magnitude away from those in real stars \citep[see review by][]{brun2017}.
A number of tracers quantifying solar activity have been used
given by different observables that are linked to different aspects of
solar activity and to different heights in the solar atmosphere.
These indices are usually highly correlated to each other due to the predominant 
eleven-year cycle,
apart from small differences on short- or long-term scale
\citep[see review by][]{usoskin2017}.

The frequencies of the acoustic modes, that propagate inside the Sun,
vary with the solar activity cycle 
\citep[][and references within]{woodard1985, libbrecht1990, chaplin2001}.
Although the variation in the mode frequencies is small 
($\la$ 0.1\%), 
it is well correlated with several solar indices 
\citep[see, for example,][and references within]{jain2009,broomhall2015}.
In fact, the seismic proxy for magnetic activity 
has been used 
to reveal an activity cycle similar to that of the Sun on other stars
\citep[][and references within]{garcia2010,kiefer2017,santos2018}.
Despite the progress in understanding the source of mode frequency
variation with activity
\citep[][among others]{gough1990,goldreich1991,li2003,dziembowski2005,mullan2007},
the picture is not completely clear. 
This work adds to clarify this
as it contributes to quantifying the observed behaviour of 
mode frequency variation with solar activity.
Here it is shown that 
the frequency increase with solar activity is not linear, but it is better
represented by a sigmoid function, 
where the frequency increase reaches a maximum.

\section{Data}

There are three sets of helioseismic data available that cover at least a broad part of 
a solar cycle and are not restricted to very low $l$ modes. I analyse all three.

The Michelson Doppler Imager (MDI)
onboard the Solar and Heliospheric Observatory (SOHO)
has a 1024 $\times$ 1024 pixel CCD with a dopplergram 
obtained every 60 seconds \citep{1995scherrer}. 
In the Medium-$l$ Program (also called the Structure Program), the image is subsampled to a 
resolution that is only one-fifth of the full-disk data and 
cropped to 90\% of the average solar radius. 
Mode parameters for 72-day long time series are provided by the
global helioseismology pipeline \citep[see][for details]{2018larson_schou}.
The data was obtained almost continuously from May 1996 to April 2011.

Helioseismic and Magnetic Imager (HMI), designed to succeeded MDI,
was launched onboard the Solar Dynamics Observatory (SDO) in February 2010
\citep{2012schou}.
It has a 4096 $\times$ 4096 pixel CCD and observes a dopplergram every 45 seconds.
HMI observes the Doppler shift of Fe I 6173 {\AA} spectral line, 
while MDI observed the Ni I 6768 {\AA} line. 
HMI samples the acoustic modes at a slightly lower height in the solar atmosphere than MDI
\citep{2011fleck}.
The mode parameters for 72-day long time series are provided by the
global helioseismology pipeline,
see \citet{2018larson_schou} for details and for a comparison with MDI.
I analysed data obtained from April 2010 to June 2017.

The Global Oscillation Network Group (GONG), 
consisting of six instruments around the world, 
observes the Doppler shift at the same line and at the same sampling time as MDI
\citep{harvey1996}.
Its 256 $\times$ 256 pixel CCD was replaced by a 1024 $\times$ 1024 one in 2001.
The mode parameters for 36-day long time series are provided by the
GONG pipeline \citep{hill1996}.
I use the central frequency obtained using Clebsch-Gordon coefficients 
after fitting each individual mode 
(radial order $n$, spherical degree $l$ and azimuthal order $m$).
In the MDI and HMI pipeline, the multiplet (2$l$ + 1 $m$'s) are fitted together
\citep{2015larson_schou}.
Unfortunately, the six ground stations are not enough to get a duty cycle
as high as MDI and HMI space missions.
HMI has a duty cycle larger than 0.96 \citep{2018larson_schou}. 
GONG's duty cycle varies from 0.67 to 0.96, 
where 98\% is lower than 0.95
\citep{2018kiefer},
while 
only 24\% of MDI's are smaller than 0.95
\citep{2015larson_schou}. 
For comparison purposes, 
only GONG data at the same time period as MDI is used.

\begin{figure}
        \includegraphics[width=\columnwidth]{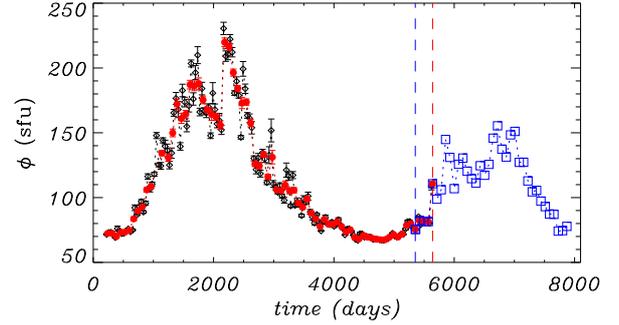}
    \caption{Radio flux 
	averaged over the analysed time series 
	(72-day for MDI and HMI and 36-day for GONG
	in black, blue and red respectively).
	The time is in Julian Day - 2450000 and covers most of solar cycle 23 and 24.
	The error bars represent the error of the mean. 
	For HMI, they are equal to or smaller than the size of the square.
	The blue and red dashed vertical lines corresponds to the first date of HMI and last of MDI
	respectively, showing the overlapping between the data sets.
}
    \label{fig_radioflux}
\end{figure}

As a physical indicator of the solar activity, 
I will use 
the daily measurements of the integrated emission from the solar disc at 2.8 GHz 
(corresponding to a 10.7 cm wavelength) 
provided by the National Resources Canada (NRC) Space Weather
\citep{covington1969,tapping1987}.
%
The solar flux density (represented in this work as $\phi$)
is measured in solar flux units (sfu), where 1 sfu = $10^{-22}W \cdot m^{-2} \cdot Hz^{-1}$.
The solar radio flux has contributions from three sources:
active regions cores, 
in and around plages and outside active regions,
and a background from the quiet Sun 
\citep{tapping1987}.
The quiet Sun radio flux is usually at 65 $\pm$ 2 sfu
\citep[and references within]{schonfeld2015}.
I use the adjusted value 
corrected for variations in the Earth-Sun distance.
The daily values are averaged over each instrument time series 
(72-day for MDI and HMI, and 36-day for GONG).
Figure~\ref{fig_radioflux} shows the averaged radio flux during MDI, GONG, and HMI observations.
Cycle 24 is approximately 0.7 smaller than Cycle 23.

\section{Analysis}

\begin{figure}
        \includegraphics[width=\columnwidth]{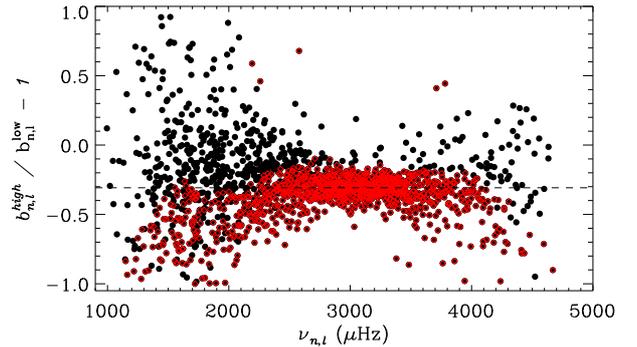}
    \caption{The figure shows the relative difference between the slope of 
a linear fit to each mode frequency observed by MDI as a function of radio flux 
using only the lower half and higher half of the radio flux. 
Modes which have a relative difference higher than 1.5$\sigma$, 
		which correspond to 65\% of the modes are shown in red.
}
    \label{fig_slope}
\end{figure}

\begin{figure}
        \includegraphics[width=\columnwidth]{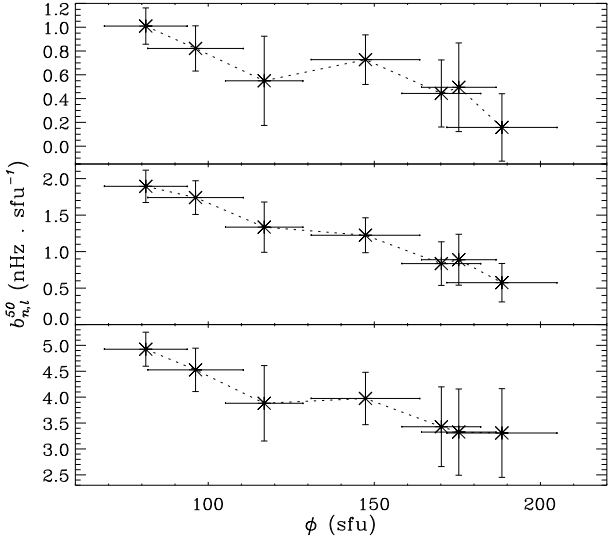}
    \caption{Variation of the slope of the frequency variation with radio flux 
observed by MDI averaged
over intervals of fifty-sfu for three different modes. 
The horizontal and vertical error bars are the standard deviation of the mean radio flux
and the slope, respectively.
From top to bottom panel, the modes are:
$n$=0, $l$=247, $\nu$ = 1583 $\mu$Hz;
$n$=3, $l$=112, $\nu$ = 2240 $\mu$Hz;
$n$=13, $l$=39, $\nu$ = 3330 $\mu$Hz.
The slope increases with mode frequency (from top to bottom panel).
}
    \label{fig_slope_variation}
\end{figure}

There is a very high linear correlation of the mode frequency variation with the solar activity 
where Pearson's linear coefficient is very close to one
\citep{jain2009}.
However, in my previous work, I found
evidence of a quadratic relation indicating a saturation effect at high solar activity
\citep{2011rabello}.
In fact, Figure~\ref{fig_slope} shows that the slope decreases as the activity increases.
I fitted a straight line to the individual mode frequencies 
observed by MDI 
with radio fluxes smaller than their mid-point (`low') and 
with fluxes larger than their mid-point (`high').
In the figure, it is plotted the relative difference between these two slopes.
%
%
The relative difference is larger than 1.5$\sigma$ for 65\% of the modes (marked in red).
The relative difference weighted mean is -0.308 $\pm$ 0.002 (dashed line).
These modes are distributed around 2900 $\mu$Hz with a FWHM of 1000 $\mu$Hz.
Using HMI data, I get similar results as MDI,
despite the smaller range in the observed radio flux,
but considering the better quality of data than MDI.
For the GONG data, 
the absolute relative difference weighted mean is 54\% larger than MDI (-0.473 $\pm$ 0.002),
and the relative difference is larger than 1.5$\sigma$ for 80\% of the modes.
Although I am using GONG data for the same time period as MDI, GONG frequencies are
obtained every 36 days, which is half of MDI time series, thus there are 
twice as many data points than MDI.
The slope decrease with solar activity is also shown in Figure~\ref{fig_slope_variation} for
three individual modes.
%
Linear regression of individual mode frequencies is performed
over subsets of fifty-sfu radio-flux intervals, 
and each subset is displaced by twenty-sfu from the previous one.

\begin{figure}
        \includegraphics[width=\columnwidth]{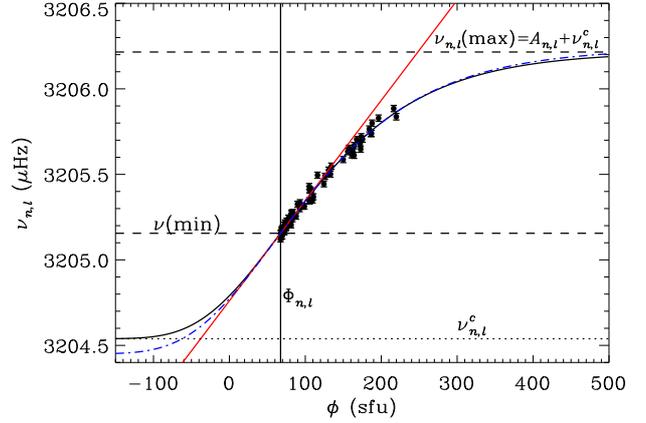}
    \caption{
	Example of the fitted model to the observed frequencies for a given mode:
	$l$=85, $n$=8 and $\nu$=3205.39 $\mu$Hz
	(black sigmoid).
	The red line indicates the maximum absolute growth rate $K_{n,l}$ at 
	the inflection point:
	$\Phi_{n,l}$ = 67.4 sfu (vertical full line).
	The upper and lower horizontal dashed lines show 
	the upper asymptote plus the offset term, $\nu_{n,l}^{\text{max}}$, and
	the frequency corresponding to the smallest observed radio flux,
	$\nu_{n,l}^{\text{min}}$, respectively. 
	The horizontal dotted line represents the constant term, $\nu^c_{n,l}$.
	The blue dot-dashed sigmoid is the same as the black one but using $\Phi_{n,l}$ = 59 sfu.
}
    \label{fig_gompertz}
\end{figure}

I propose a Gompertz model to represent the frequency variation with solar activity
instead of a linear fit.
The Gompertz model is a frequently used sigmoid function, especially in biological and social sciences.
It has been often used to model biological ageing or
%
the growth of plants, animals, bacteria and even tumour,
but it has also been used on other research fields such as 
Business, Computer Science, Engineering, Geophysics, and Physics. 
The growth rate is quick at first, but eventually, it slows down and then levels off.
%
%
Numerous parametrization 
of the Gompertz model can be found in the literature.
I use here the one given by \citet[equation~18]{tjorve2017},
where each parameter only affects one shape characteristic. 
The Gompertz model plus a constant term ($\nu^c_{n,l}$)
was fitted
to each individual mode frequency: $\nu_{n,l}(\phi)$ 
where $\phi$ is the mean radio flux at each time interval.
%
\begin{equation}
\nu_{n,l}(\phi) = A_{n,l} \exp{\{-\exp{[-e k_{n,l} (\phi - \Phi_{n,l})}]\}} + \nu^c_{n,l}
\label{eq_gompertz}
\end{equation}
where 
$\exp{(x)}=e^x$ and $e$ is the base of the natural logarithm $\ln$.
$A_{n,l}$ represents the upper asymptote, 
$\Phi_{n,l}$ is the radio flux at the inflection point, and
$k_{n,l}$ is the maximum relative growth rate (Figure~\ref{fig_gompertz}).
The maximum absolute growth rate is given by $K_{n,l} = k_{n,l} \cdot A_{n,l}$
and it is equal to the tangent at the inflection point.
The value at inflection is locked at 36.8\% of the upper asymptote:
\begin{equation}
\nu_{n,l}(\Phi_{n,l}) = A_{n,l} e^{-1} + \nu^c_{n,l}.
\label{eq_inflection}
\end{equation}

The proposed Gompertz function has a physical meaning only for $\phi$ 
equal to or larger than the quiet Sun radio flux.
In equation~\ref{eq_gompertz},
when $\phi \rightarrow -\infty$, $\nu_{n,l} \rightarrow \nu^c_{n,l}$.
The constant term, $\nu^c_{n,l}$, does not have a physical meaning.
From the model, one can estimate the maximum possible frequency 
for each mode, $\nu_{n,l}^\text{max}$,
when $\phi \rightarrow +\infty$, $\nu_{n,l} \rightarrow \nu_{n,l}^\text{max} = A_{n,l} + \nu^c_{n,l}$.

The frequency shift rate decreases as the radio flux increases
(Figure~\ref{fig_slope_variation}), thus
the inflection point, $\Phi_{n,l}$, is at the minimum 
observed solar radio flux or at an even smaller value.
I assume
that $\Phi_{n,l}$ is the same for all modes and equal to the minimum observed solar flux
in the data ($\phi^\text{min}$=67.4 sfu for MDI and GONG).
In this case, 
the minimum-to-maximum frequency shift 
is given by:
%
\begin{equation}
\Delta\nu_{n,l} = \nu_{n,l}^\text{max} - \nu_{n,l}^\text{min} = A_{n,l} (1 - e^{-1}),
\label{eq_maxmin}
\end{equation}
where $\nu_{n,l}^\text{min}$ is the mode frequency at $\phi = \Phi_{n,l}$.
The  $\Phi_{n,l}$ parameter shifts the sigmoid curve horizontally 
without changing its shape \citep{tjorve2017}.

Equation~\ref{eq_gompertz} was fitted using a Levenberg-Marquardt least-squares fit
\citep{2012markwardt}
with only two free parameters $k_{n,l}$ and $\nu^c_{n,l}$.
$\Phi_{n,l}$ is fixed at 67.4 sfu. 
Applying Equation~\ref{eq_inflection}, we have
$A_{n,l} = [\nu_{n,l}^\text{min} - \nu^c_{n,l}] e$,
where $\nu_{n,l}(\Phi_{n,l}) = \nu_{n,l}(\phi^\text{min})=\nu_{n,l}^\text{min}$.
After doing a linear regression
to one-third of the smallest values of the radio flux,
$\nu_{n,l}^\text{min}$ is given by
the fitted frequency corresponding to the smallest observed radio flux.
Out of 1868 modes observed by MDI, 
1605 were successfully fitted indicating a success rate of 86\%.
The success rate is 83\% and 77\% for GONG and HMI respectively.
HMI observations were taken during solar cycle 24 which has 
a smaller maximum (155 sfu) than MDI and GONG observations (220 sfu).
Also, the used HMI data has a higher minimum (74 sfu)
and 
they are not as long as MDI, 
there are 74 time series in MDI data against only 37 for HMI.
The same value was used for HMI as for MDI and GONG for $\Phi_{n,l}$ = 67.4 sfu.
The weaker Cycle 24 makes the frequency shift smaller and
worsens the Gompertz fitting, specially for
low-frequency modes ($\nu$ < 2.5 mHz)
that have a small frequency-shift gradient.

\section{Results and Discussion using MDI data}

\begin{figure}
        \includegraphics[width=\columnwidth]{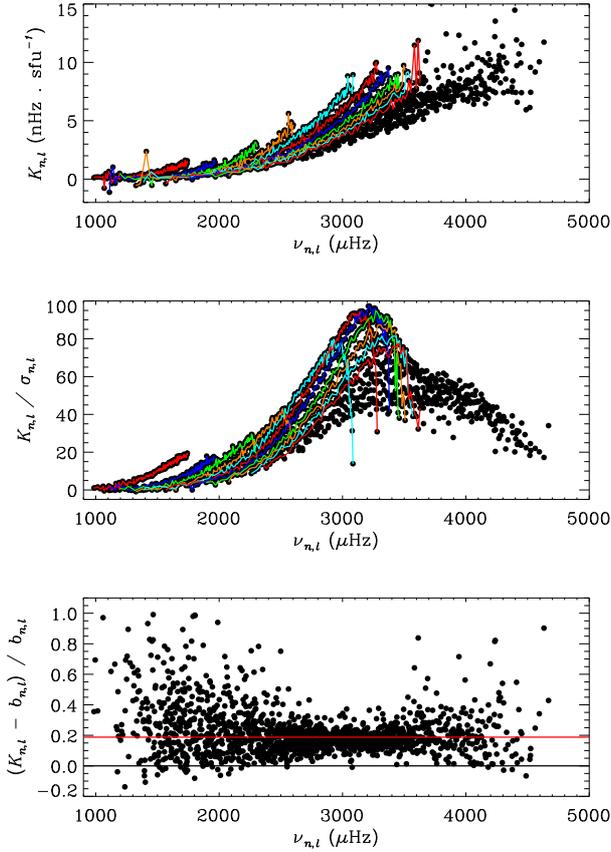}
    \caption{
Top Panel. Fitted values of the maximum absolute growth rate $K_{n,l}$.
Middle Panel. Fitted values of $K_{n,l}$ divided by its fitting uncertainty ($\sigma_{n,l}$).
Bottom Panel. Comparison between $K_{n,l}$ and the linear fitting slope,
$b_{n,l}$.
The coloured lines, in the top and middle panels, 
show the order of the modes from $n$=0 until 10 
repeating the sequence in the order red, blue, green, salmon, and cyan.
The red horizontal line in the bottom panel represents the median of the relative differences.
}
    \label{fig_ku}
\end{figure}

Figure~\ref{fig_ku} (top panel) shows the maximum absolute growth rate, $K_{n,l}$, 
which increases with frequency
and decreases with mode order $n$. 
The fitted values are several times its fitting uncertainty ($\sigma_{n,l}$) 
as shown in the middle panel,
except for 36\% of modes which are smaller than 5$\sigma$.
These modes have frequencies smaller than 2000 $\mu$Hz.
The bottom panel compares $K_{n,l}$ with 
the slope of a linear fit ($b_{n,l}$). 
The red horizontal line shows the median of the relative difference
equal to 0.19.
%
$K_{n,l}$
is closer to the slope fitting only radio fluxes smaller than its middle point 
($b_{n,l}^{\text{low}}$ in Figure~\ref{fig_slope}).
In this case, the median is equal to 0.04.

\begin{figure}
        \includegraphics[width=\columnwidth]{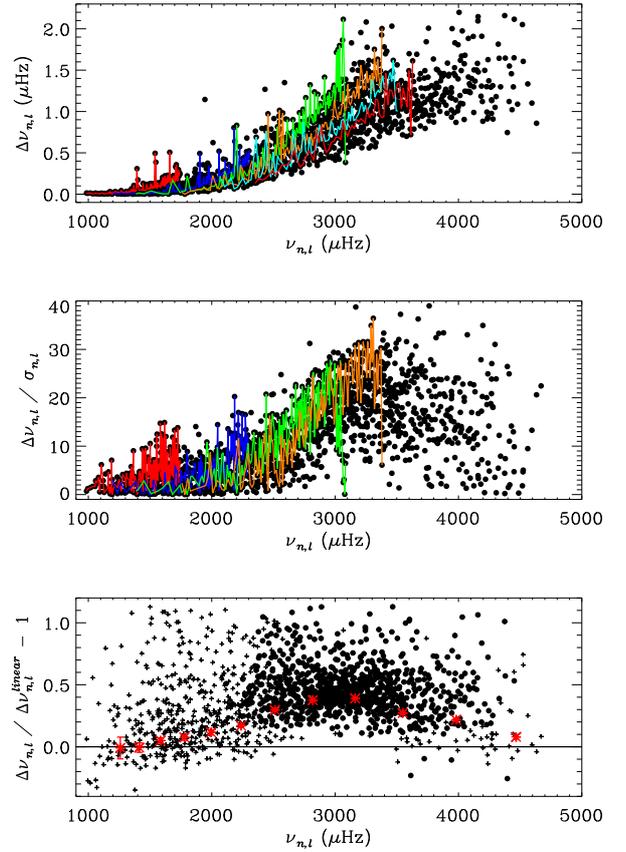}
    \caption{
Top Panel. Minimum-to-maximum frequency shift, $\Delta\nu_{n,l}$ (Equation~\ref{eq_maxmin}).
Middle Panel. $\Delta\nu_{n,l}$ divided by its uncertainty ($\sigma_{n,l}$).
Bottom Panel. Relative difference between $\Delta\nu_{n,l}$ and 
the linear frequency shift: 
$\Delta\nu_{n,l}^{\text{linear}} = b_{n,l} \cdot \Delta\phi$,
where $\Delta\phi$ = 145 sfu.
The modes which have a relative difference larger than 2$\sigma$ 
are represented with circles instead of crosses.
The red stars are weighted averages of 0.05 intervals in log$\nu_{n,l}$ (in $\mu$Hz) and
their error bars are the error of the mean which is smaller than the symbol except for the first two stars.
The coloured lines, in the top and middle panels, 
show the order of the modes from $n$ = 0, 2, 4, 6, 8, and 10 which are
red, blue, green, salmon, cyan, and red respectively for the top panel 
and only until $n$=6 for the middle panel. 
}
    \label{fig_deltanu}
\end{figure}

Figure~\ref{fig_deltanu} shows the minimum-to-maximum frequency shift
(Equation~\ref{eq_maxmin})
which also increases with frequency and decreases with order $n$ (top panel).
There is scatter in its variation with $n$.
The fitted values are several times its sigma as shown in the middle panel,
except for 28\% of all modes which are smaller than 5$\sigma$.
%
The bottom panel shows the relative difference between $\Delta\nu_{n,l}$
and 
frequency shift obtained by the linear fit defined as
$\Delta\nu_{n,l}^{\text{linear}} = b_{n,l} \cdot \Delta\phi$,
where $\Delta\phi$ = ($\phi^{\text{max}} - \phi^{\text{min}}$) = 145 sfu
was chosen to get similar values to $\Delta\nu_{n,l}$ at lower frequencies.
The observed radio-flux difference, 
$\Delta\phi$,
for MDI is 220 sfu.
The modes which have a relative difference larger than 2$\sigma$ are represented with circles instead of crosses.
$\Delta\nu_{n,l}$ gives the maximum frequency shift possible,
while $\Delta\nu_{n,l}^{\text{linear}}$ depends on the cycle that is being analysed.
Although their absolute differences are very small 
($\la$ 0.2 $\mu$Hz), 
there is a systematic variation in the relative difference with frequency
(Figure~\ref{fig_deltanu}).

\begin{figure}
        \includegraphics[width=\columnwidth]{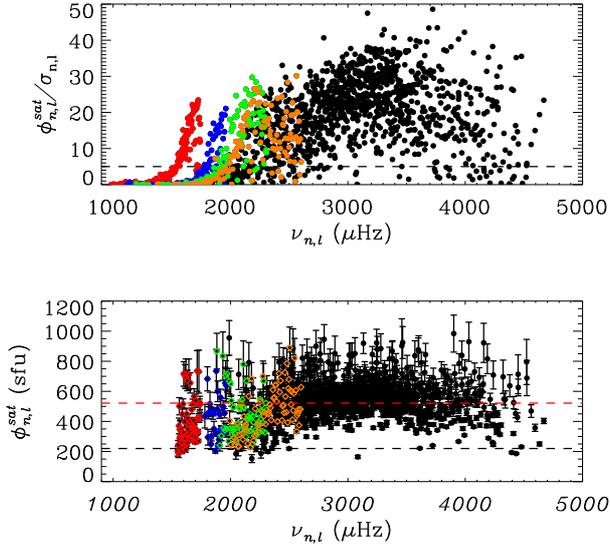}
    \caption{
The mode saturation $\phi_{n,l}^\text{sat}$ given by Equation~\ref{eq_sat}.
Top Panel. Mode saturation divided by its uncertainty ($\sigma_{n,l}$).
The horizontal dashed line is at 5$\sigma$.
Bottom Panel. Only modes with saturation larger than 5$\sigma$. 
The red and black horizontal dashed lines represents the median saturation
and the maximum observed radio flux by MDI and GONG.
The coloured symbols, in both panels,
show the order of the modes only from $n$=0 until 3 which are
red, blue, green, salmon respectively. 
}
    \label{fig_saturation}
\end{figure}

The saturation is defined as the value of radio flux, $\phi_{n,l}^\text{sat}$,
whose corresponding mode frequency differs from $\nu_{n,l}^\text{max}$
by less than 0.0004\% (i.e., $\la$ 0.01 $\mu$Hz in 2500 $\mu$Hz).
From Equation~\ref{eq_gompertz}, we have:
\begin{equation}
\phi_{n,l}^{\text{sat}} = \Phi_{n,l} - 
\ln \left[ -\ln \left( 1 - \frac{4 \times 10^{-6} \cdot \nu_{n,l}}{A_{n,l}} \right) \right]
\cdot \frac{A_{n,l}}{e \cdot k_{n,l}}.
\label{eq_sat}
\end{equation}
In Figure~\ref{fig_saturation} (top panel), 
the saturation is larger than 5$\sigma$ for 77\% of the modes,
where the saturation uncertainty is calculated by 
propagation of the fitting uncertainties.
The bottom panel shows only well-determined saturation (i.e., larger than 5$\sigma$).
Their median saturation is 522 sfu 
(horizontal red dashed line)
and 70\% of the modes are within (522$\pm$140) sfu.
Their weighted mean  
is (428.7 $\pm$ 0.6) sfu.
The largest radio flux, averaged over a 72-day, during Cycle 23 is 220 sfu
(horizontal black dashed line).
Half of the modes will reach saturation only if a cycle is 
more than two times stronger than Cycle 23.
Cycle 19, the strongest cycle in the last century,
during which there was 72-day radio flux average as large as 273 sfu, is only 24\% stronger than Cycle 23.

The deviation from a linear fit is very small.
The significance of the results is tested by adding 
normally distributed random numbers with a standard deviation equal to 
the observed frequency uncertainty,
$e_{n,l}(\phi)$,
to the linear fitted mode frequency,
$\nu_{n,l}^{\text{linear}}(\phi) = a_{n,l} + b_{n,l} \cdot \phi$,
and fitting the Gompertz model 
to each individual mode frequency for 1000 realizations,
in the same way as the observations.
The smallest median saturation over all modes, 
which have $\phi_{n,l}^\text{sat} > 5\sigma$,
is 638 sfu, which is 23\% larger than the observed one;
and only 1\% of the realizations have a median smaller than 680 sfu 
(i.e., 30\% larger than the observed one).
Thus, the results from the Gompertz model using MDI data are statistically significant.

As mentioned, it is possible that $\Phi_{n,l}$ is smaller than the minimum observed radio flux.
Assuming a value for $\Phi_{n,l}$ that is 8 sfu smaller than the minimum observed by MDI
(12\% smaller),
the Gompertz fitted parameters change by a small amount.
The mean saturation is 10.9 sfu larger than before (2.5\%)
and the median increases by 29 sfu (5.6\%).
Figure~\ref{fig_gompertz} shows the sigmoid for a given mode
with the smaller $\Phi_{n,l}$ in blue 
to be compared with the previous one in black.
Since 1947, 
the lowest observed daily radio flux is equal to 62 sfu, 
which happened during the 1953-1954 minimum.
During the recent unusually low 2007-2008 minimum, the smallest daily value is 65 sfu.

\begin{figure}
        \includegraphics[width=\columnwidth]{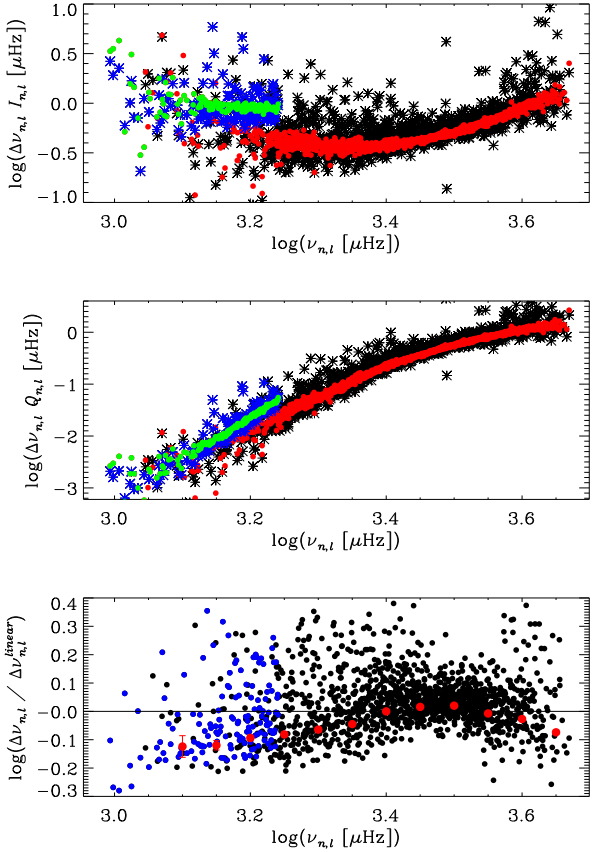}
    \caption{ 
The frequency shift, $\Delta\nu_{n,l}$, 
scaled by $I_{n,l}$ (top panel)
and by the normalized mode inertia $Q_{n,l}$ (middle panel),
expressed in base-10 logarithm.
The $p$ ($f$) modes are in black and red (blue and green) for 
the Gompertz frequency shift (Equation~\ref{eq_maxmin})
and 
the linear frequency shift, $\Delta\nu_{n,l}^{\text{linear}}$ 
(using $\Delta\phi$ = 195 sfu), respectively.
The logarithm of the ratio between Gompertz and linear frequency shift
is plotted in the bottom panel, 
where 
$f$ and $p$ modes are the blue and black small circles respectively.
The red circles are weighted averages in 0.05-$\log{\nu}$ $\mu$Hz intervals.
}
    \label{fig_deltanuqnl}
\end{figure}

The frequency shifts depend mostly on the mode frequency after they have been weighted
by the mode inertia, 
$\Delta\nu_{n,l} \propto \nu^{\alpha}_{n,l} / I_{n,l}$,
indicating that the source of the perturbations is close to the solar surface
\citep{libbrecht1990}.
Figure~\ref{fig_deltanuqnl} shows $\Delta\nu_{n,l}$ scaled by $I_{n,l}$ (top panel)
and by the normalized mode inertia $Q_{n,l}$ (middle panel),
expressed in base-10 logarithm.
$Q_{n,l}$ is 
normalized by the inertia of a radial mode of the same frequency \citep{christensen1989}.
The mode inertia was obtained using model `S' \citep{christensen1996}.
The linear frequency shift, $\Delta\nu_{n,l}^{\text{linear}}$, 
is also plotted for comparison
and 
it is the same as defined in Figure~\ref{fig_deltanu}, except that here
$\Delta\phi$ = 195 sfu
was chosen to get similar values to $\Delta\nu_{n,l}$ at all frequencies.
The $p$ modes are in black and red for the Gompertz fitting (Equation~\ref{eq_maxmin})
and linear fitting, respectively.
The $f$ modes are in blue and green respectively
and 
have different behaviour from $p$ modes for both fitting methods. 
The response of the frequency shift scaled by the mode inertia
depends upon the physical mechanism responsible for changing 
the mode frequencies during the solar cycle \citep{gough1990}.
Thus, 
although both $f$ and $p$ modes are well correlated with radio flux, 
the cause of frequency variation with solar activity are different
\citep[see, for example,][]{dziembowski2005}.

In Figure~\ref{fig_deltanuqnl},
a straight line fitting to
$\log(\delta\nu_{n,l} \cdot Q_{n,l})$ in the middle panel
corresponds to a parabolic-like shape in the top panel
\citep[][Figure 6]{rabellosoares2008adv},
hence $\alpha$ varies with frequency.
A linear regression to 
$\log(\Delta\nu_{n,l}^{\text{linear}} \cdot Q_{n,l})$
shows a clear change in the slope 
around 2500 $\mu$Hz
and the slopes,
$\gamma$, are:
\begin{equation} \label{eq1}
\begin{split}
\gamma = 5.97 \pm 0.02 & \quad \text{for} \quad \nu < 2500 \; \mu Hz \\
\gamma = 3.57 \pm 0.01 & \quad \text{for} \quad \nu > 2500 \; \mu Hz.
\end{split}
\end{equation}
for $p$ modes. The $f$ modes have a larger coefficient, $\gamma$ = 7.90 $\pm$ 0.06.
As pointed out by \citet{chaplin2001}, 
the upper turning point of the modes gets deeper as the mode frequency decreases
and, around 2500$\mu$Hz,
the gradient becomes suddenly an order of magnitude larger than before 
and
low-frequency modes are reflected back at much deeper layers of the solar atmosphere,
which might affect the frequency shift since it is expected 
that the source of perturbation is close to the surface.
Although there is a good agreement between both frequency shifts, 
the logarithm of the ratio between Gompertz and linear frequency shift
is plotted in the bottom panel, 
where 
$f$ and $p$ modes are the blue and black small circles respectively.
The red circles are weighted averages in 0.05-$\log{\nu}$ $\mu$Hz intervals.
They show a clear difference between the two fitting methods.
Fitting a straight line to $\log{ \Delta\nu_{n,l}/\Delta\nu_{n,l}^{\text{linear}} }$,
one gets $\Delta\gamma = \gamma_{\text{Gompertz}} - \gamma_{\text{linear}}$:
\begin{equation} \label{eq2}
\begin{split}
\Delta\gamma & = 0.40 \pm 0.04 \quad \text{for} \quad \nu < 2500 \; \mu Hz \\
\Delta\gamma & = -0.31 \pm 0.02 \quad \text{for} \quad \nu > 2500 \; \mu Hz.
\end{split}
\end{equation}
The Gompertz fitting estimates a slightly different behaviour
for the scaled frequency shift than the linear fitting,
where 
$\gamma$ is larger by 7\% and smaller by 9\%
for frequencies smaller and larger than 2500 $\mu$Hz respectively.

\section{Results and Discussion using HMI and GONG data}

{\bf GONG.} 
The saturation weighted mean,
calculated using Equation~\ref{eq_sat}, 
is (395.4 $\pm$ 0.3) sfu, which is similar to the one obtained using MDI data,
it is 8\% smaller than MDI.
The median saturation, calculated using modes with saturation larger than
5$\sigma$ (which corresponds to 86\% of the fitted modes), is
434 sfu and 17\% smaller than MDI.
Similarly to MDI, 
I fitted the Gompertz model to
the linear fitted GONG mode frequencies after adding
normally distributed random numbers with a standard deviation equal to 
the uncertainty of the GONG observed frequencies to test the significance of the results.
For 1000 realizations,
the smallest median saturation is 858 sfu, using only modes with 
saturation larger than 5$\sigma$, which is almost twice the median 
of the observations.
The results from the Gompertz model using GONG data
are even more robust than for MDI.
For GONG, there are twice as many points as for MDI during the same time period due
to the shorter time series (36-day) that the frequencies were obtained.

{\bf HMI}. 
The saturation weighted mean, calculated using Equation~\ref{eq_sat} 
with $\Phi_{n,l}$=67.4 sfu same as for MDI,
is (226.7 $\pm$ 0.3) sfu, which is almost half the saturation obtained using MDI data.
The median saturation, calculated using modes with saturation larger than
5$\sigma$ (which corresponds to 74\% of the fitted modes), is 281 sfu.
There is no significant change in the saturation determination for HMI
using $\Phi_{n,l}$ = 74 sfu, 
which is observed minimum radio flux for the HMI data used here.
Repeating the Monte Carlo simulation for HMI data, 
using only modes with saturation larger than 5$\sigma$, 
the smallest median saturation is 200 sfu, which smaller than the observed median.
There are 18 realizations with a median smaller than the observed one
and 85 with a weighted mean smaller than the observed one.
The Gompertz fitting using HMI data is not statistically significant.
The results are due to the randomness introduced by observed frequency uncertainties.
Cycle 24 is among the weakest cycles in the last 200 years,
As the frequency shift is far from saturation, it is not possible to determine it.
The saturation obtained for MDI and GONG could be a lower limit
for the true saturation.
A stronger cycle is needed to get a better determination.

\section{Conclusions}

It is shown that the rate at which the frequency increases with the solar radio flux 
decreases as the solar radio flux increases using MDI, HMI and GONG data.
A sigmoid function is proposed instead of a linear fit 
to the frequency dependence with solar radio flux:
the Gompertz model as defined by \citet{tjorve2017} plus a constant
(Equation~\ref{eq_gompertz}).

Although the deviation from a linear fit is small, 
it is shown that the sigmoid fitting is statistically significant 
using MDI and GONG data. 
This is not the case for HMI data, which was 
obtained during solar Cycle 24, a much weaker cycle than the previous 
one when MDI and GONG data were taken.
Using MDI and GONG data, 
I estimate a saturation level of four-hundred sfu averaging over all modes,
which is 50\% larger than Cycle 19, the strongest cycle in the last century.
The saturation observed by GONG is a few percents lower than by MDI.
The exact mechanism causing the frequencies to increase with solar activity is not precisely known, which makes it difficult to speculate on 
why they increase at a smaller rate at high activity levels indicating a possible saturation at a high enough activity level.

The minimum-to-maximum frequency difference 
scaled by the normalized mode inertia
obtained here has different behaviour than
the commonly used difference between the frequencies obtained at or close to the cycle maximum activity and to its minimum (i.e., assuming a linear fit).
The slope is 7\% larger and 9\% smaller than assuming a linear fit
for with frequency smaller and larger than 2500 $\mu$Hz, respectively.
This dissimilarity in the response is a direct 
consequence of the nature of the perturbation causing the frequency variation
and, although small, it should not be neglected.

\section*{Acknowledgements}

This research was supported in part by Minas Gerais State Agency for Research and Development (FAPEMIG), Brazil.

Solar radio flux data is provided by the National Resources Canada (NRC) Space Weather.
MDI data is 
provided by the SOHO/MDI consortium. SOHO is a project of international cooperation between ESA and NASA.
HMI data is provided by NASA/SDO and HMI science team.
This work utilizes data obtained by the Global Oscillation Network
Group (GONG) program, managed by the National Solar Observatory, which
is operated by AURA, Inc. under a cooperative agreement with the
National Science Foundation. The data were acquired by instruments
operated by the Big Bear Solar Observatory, High Altitude Observatory,
Learmonth Solar Observatory, Udaipur Solar Observatory, Instituto de
Astrof\'{\i}sica de Canarias, and Cerro Tololo Interamerican
Observatory.



\bibliographystyle{mnras}
\bibliography{rabellosoares} 

\begin{thebibliography}{}
\makeatletter
\relax
\def\mn@urlcharsother{\let\do\@makeother \do\$\do\&\do\#\do\^\do\_\do\%\do\~}
\def\mn@doi{\begingroup\mn@urlcharsother \@ifnextchar [ {\mn@doi@}
  {\mn@doi@[]}}
\def\mn@doi@[#1]#2{\def\@tempa{#1}\ifx\@tempa\@empty \href
  {http://dx.doi.org/#2} {doi:#2}\else \href {http://dx.doi.org/#2} {#1}\fi
  \endgroup}
\def\mn@eprint#1#2{\mn@eprint@#1:#2::\@nil}
\def\mn@eprint@arXiv#1{\href {http://arxiv.org/abs/#1} {{\tt arXiv:#1}}}
\def\mn@eprint@dblp#1{\href {http://dblp.uni-trier.de/rec/bibtex/#1.xml}
  {dblp:#1}}
\def\mn@eprint@#1:#2:#3:#4\@nil{\def\@tempa {#1}\def\@tempb {#2}\def\@tempc
  {#3}\ifx \@tempc \@empty \let \@tempc \@tempb \let \@tempb \@tempa \fi \ifx
  \@tempb \@empty \def\@tempb {arXiv}\fi \@ifundefined
  {mn@eprint@\@tempb}{\@tempb:\@tempc}{\expandafter \expandafter \csname
  mn@eprint@\@tempb\endcsname \expandafter{\@tempc}}}

\bibitem[\protect\citeauthoryear{{Broomhall} \& {Nakariakov}}{{Broomhall} \&
  {Nakariakov}}{2015}]{broomhall2015}
{Broomhall} A.-M.,  {Nakariakov} V.~M.,  2015, \mn@doi [\solphys]
  {10.1007/s11207-015-0728-6}, \href
  {http://adsabs.harvard.edu/abs/2015SoPh..290.3095B} {290, 3095}

\bibitem[\protect\citeauthoryear{{Brun} \& {Browning}}{{Brun} \&
  {Browning}}{2017}]{brun2017}
{Brun} A.~S.,  {Browning} M.~K.,  2017, \mn@doi [Living Reviews in Solar
  Physics] {10.1007/s41116-017-0007-8}, \href
  {http://adsabs.harvard.edu/abs/2017LRSP...14....4B} {14, 4}

\bibitem[\protect\citeauthoryear{{Chaplin}, {Appourchaux}, {Elsworth}, {Isaak}
  \& {New}}{{Chaplin} et~al.}{2001}]{chaplin2001}
{Chaplin} W.~J.,  {Appourchaux} T.,  {Elsworth} Y.,  {Isaak} G.~R.,   {New} R.,
   2001, \mn@doi [\mnras] {10.1046/j.1365-8711.2001.04357.x}, \href
  {http://adsabs.harvard.edu/abs/2001MNRAS.324..910C} {324, 910}

\bibitem[\protect\citeauthoryear{{Christensen-Dalsgaard}, {Thompson}  \&
  {Gough}}{{Christensen-Dalsgaard} et~al.}{1989}]{christensen1989}
{Christensen-Dalsgaard} J.,  {Thompson} M.~J.,   {Gough} D.~O.,  1989, \mn@doi
  [\mnras] {10.1093/mnras/238.2.481}, \href
  {http://adsabs.harvard.edu/abs/1989MNRAS.238..481C} {238, 481}

\bibitem[\protect\citeauthoryear{{Christensen-Dalsgaard}
  et~al.,}{{Christensen-Dalsgaard} et~al.}{1996}]{christensen1996}
{Christensen-Dalsgaard} J.,  et~al., 1996, \mn@doi [Science]
  {10.1126/science.272.5266.1286}, \href
  {http://adsabs.harvard.edu/abs/1996Sci...272.1286C} {272, 1286}

\bibitem[\protect\citeauthoryear{{Covington}}{{Covington}}{1969}]{covington1969}
{Covington} A.~E.,  1969, \jrasc, \href
  {http://adsabs.harvard.edu/abs/1969JRASC..63..125C} {63, 125}

\bibitem[\protect\citeauthoryear{{Dziembowski} \& {Goode}}{{Dziembowski} \&
  {Goode}}{2005}]{dziembowski2005}
{Dziembowski} W.~A.,  {Goode} P.~R.,  2005, \mn@doi [\apj] {10.1086/429712},
  \href {http://adsabs.harvard.edu/abs/2005ApJ...625..548D} {625, 548}

\bibitem[\protect\citeauthoryear{{Fleck}, {Couvidat}  \& {Straus}}{{Fleck}
  et~al.}{2011}]{2011fleck}
{Fleck} B.,  {Couvidat} S.,   {Straus} T.,  2011, \mn@doi [\solphys]
  {10.1007/s11207-011-9783-9}, \href
  {http://adsabs.harvard.edu/abs/2011SoPh..271...27F} {271, 27}

\bibitem[\protect\citeauthoryear{{Garc{\'{\i}}a}, {Mathur}, {Salabert},
  {Ballot}, {R{\'e}gulo}, {Metcalfe}  \& {Baglin}}{{Garc{\'{\i}}a}
  et~al.}{2010}]{garcia2010}
{Garc{\'{\i}}a} R.~A.,  {Mathur} S.,  {Salabert} D.,  {Ballot} J.,
  {R{\'e}gulo} C.,  {Metcalfe} T.~S.,   {Baglin} A.,  2010, \mn@doi [Science]
  {10.1126/science.1191064}, \href
  {http://adsabs.harvard.edu/abs/2010Sci...329.1032G} {329, 1032}

\bibitem[\protect\citeauthoryear{{Goldreich}, {Murray}, {Willette}  \&
  {Kumar}}{{Goldreich} et~al.}{1991}]{goldreich1991}
{Goldreich} P.,  {Murray} N.,  {Willette} G.,   {Kumar} P.,  1991, \mn@doi
  [\apj] {10.1086/169858}, \href
  {http://adsabs.harvard.edu/abs/1991ApJ...370..752G} {370, 752}

\bibitem[\protect\citeauthoryear{{Gough}}{{Gough}}{1990}]{gough1990}
{Gough} D.~O.,  1990, in {Osaki} Y.,  {Shibahashi} H.,  eds,  Lecture Notes in
  Physics, Berlin Springer Verlag Vol. 367, Progress of Seismology of the Sun
  and Stars. p.~283, \mn@doi{10.1007/3-540-53091-6}

\bibitem[\protect\citeauthoryear{{Harvey} et~al.,}{{Harvey}
  et~al.}{1996}]{harvey1996}
{Harvey} J.~W.,  et~al., 1996, \mn@doi [Science]
  {10.1126/science.272.5266.1284}, \href
  {http://adsabs.harvard.edu/abs/1996Sci...272.1284H} {272, 1284}

\bibitem[\protect\citeauthoryear{{Hill} et~al.,}{{Hill}
  et~al.}{1996}]{hill1996}
{Hill} F.,  et~al., 1996, \mn@doi [Science] {10.1126/science.272.5266.1292},
  \href {http://adsabs.harvard.edu/abs/1996Sci...272.1292H} {272, 1292}

\bibitem[\protect\citeauthoryear{{Jain}, {Tripathy}  \& {Hill}}{{Jain}
  et~al.}{2009}]{jain2009}
{Jain} K.,  {Tripathy} S.~C.,   {Hill} F.,  2009, \mn@doi [\apj]
  {10.1088/0004-637X/695/2/1567}, \href
  {http://adsabs.harvard.edu/abs/2009ApJ...695.1567J} {695, 1567}

\bibitem[\protect\citeauthoryear{{Kiefer}, {Schad}, {Davies}  \&
  {Roth}}{{Kiefer} et~al.}{2017}]{kiefer2017}
{Kiefer} R.,  {Schad} A.,  {Davies} G.,   {Roth} M.,  2017, \mn@doi [\aap]
  {10.1051/0004-6361/201628469}, \href
  {http://adsabs.harvard.edu/abs/2017A%26A...598A..77K} {598, A77}

\bibitem[\protect\citeauthoryear{{Kiefer}, {Komm}, {Hill}, {Broomhall}  \&
  {Roth}}{{Kiefer} et~al.}{2018}]{2018kiefer}
{Kiefer} R.,  {Komm} R.,  {Hill} F.,  {Broomhall} A.-M.,   {Roth} M.,  2018,
  \mn@doi [\solphys] {10.1007/s11207-018-1370-x}, \href
  {http://adsabs.harvard.edu/abs/2018SoPh..293..151K} {293, 151}

\bibitem[\protect\citeauthoryear{{Larson} \& {Schou}}{{Larson} \&
  {Schou}}{2015}]{2015larson_schou}
{Larson} T.~P.,  {Schou} J.,  2015, \mn@doi [\solphys]
  {10.1007/s11207-015-0792-y}, \href
  {http://adsabs.harvard.edu/abs/2015SoPh..290.3221L} {290, 3221}

\bibitem[\protect\citeauthoryear{{Larson} \& {Schou}}{{Larson} \&
  {Schou}}{2018}]{2018larson_schou}
{Larson} T.~P.,  {Schou} J.,  2018, \mn@doi [\solphys]
  {10.1007/s11207-017-1201-5}, \href
  {http://adsabs.harvard.edu/abs/2018SoPh..293...29L} {293, 29}

\bibitem[\protect\citeauthoryear{{Li}, {Basu}, {Sofia}, {Robinson}, {Demarque}
  \& {Guenther}}{{Li} et~al.}{2003}]{li2003}
{Li} L.~H.,  {Basu} S.,  {Sofia} S.,  {Robinson} F.~J.,  {Demarque} P.,
  {Guenther} D.~B.,  2003, \mn@doi [\apj] {10.1086/375484}, \href
  {http://adsabs.harvard.edu/abs/2003ApJ...591.1267L} {591, 1267}

\bibitem[\protect\citeauthoryear{{Libbrecht} \& {Woodard}}{{Libbrecht} \&
  {Woodard}}{1990}]{libbrecht1990}
{Libbrecht} K.~G.,  {Woodard} M.~F.,  1990, \mn@doi [\nat] {10.1038/345779a0},
  \href {http://adsabs.harvard.edu/abs/1990Natur.345..779L} {345, 779}

\bibitem[\protect\citeauthoryear{{Markwardt}}{{Markwardt}}{2012}]{2012markwardt}
{Markwardt} C.,  2012, {MPFIT: Robust non-linear least squares curve fitting},
  Astrophysics Source Code Library (\mn@eprint {ascl} {1208.019})

\bibitem[\protect\citeauthoryear{{Mullan}, {MacDonald}  \& {Townsend}}{{Mullan}
  et~al.}{2007}]{mullan2007}
{Mullan} D.~J.,  {MacDonald} J.,   {Townsend} R.~H.~D.,  2007, \mn@doi [\apj]
  {10.1086/522559}, \href {http://adsabs.harvard.edu/abs/2007ApJ...670.1420M}
  {670, 1420}

\bibitem[\protect\citeauthoryear{{Rabello-Soares}}{{Rabello-Soares}}{2011}]{2011rabello}
{Rabello-Soares} M.~C.,  2011, \mn@doi [Journal of Physics Conference Series]
  {10.1088/1742-6596/271/1/012026}, \href
  {http://adsabs.harvard.edu/abs/2011JPhCS.271a2026R} {271, 012026}

\bibitem[\protect\citeauthoryear{{Rabello-Soares}, {Korzennik}  \&
  {Schou}}{{Rabello-Soares} et~al.}{2008}]{rabellosoares2008adv}
{Rabello-Soares} M.~C.,  {Korzennik} S.~G.,   {Schou} J.,  2008, \mn@doi
  [Advances in Space Research] {10.1016/j.asr.2007.03.014}, \href
  {http://adsabs.harvard.edu/abs/2008AdSpR..41..861R} {41, 861}

\bibitem[\protect\citeauthoryear{{Santos} et~al.,}{{Santos}
  et~al.}{2018}]{santos2018}
{Santos} A.~R.~G.,  et~al., 2018, \mn@doi [\apjs] {10.3847/1538-4365/aac9b6},
  \href {http://adsabs.harvard.edu/abs/2018ApJS..237...17S} {237, 17}

\bibitem[\protect\citeauthoryear{{Scherrer} et~al.,}{{Scherrer}
  et~al.}{1995}]{1995scherrer}
{Scherrer} P.~H.,  et~al., 1995, \mn@doi [\solphys] {10.1007/BF00733429}, \href
  {http://adsabs.harvard.edu/abs/1995SoPh..162..129S} {162, 129}

\bibitem[\protect\citeauthoryear{{Schonfeld}, {White}, {Henney}, {Arge}  \&
  {McAteer}}{{Schonfeld} et~al.}{2015}]{schonfeld2015}
{Schonfeld} S.~J.,  {White} S.~M.,  {Henney} C.~J.,  {Arge} C.~N.,   {McAteer}
  R.~T.~J.,  2015, \mn@doi [\apj] {10.1088/0004-637X/808/1/29}, \href
  {http://adsabs.harvard.edu/abs/2015ApJ...808...29S} {808, 29}

\bibitem[\protect\citeauthoryear{{Schou} et~al.,}{{Schou}
  et~al.}{2012}]{2012schou}
{Schou} J.,  et~al., 2012, \mn@doi [\solphys] {10.1007/s11207-011-9842-2},
  \href {http://adsabs.harvard.edu/abs/2012SoPh..275..229S} {275, 229}

\bibitem[\protect\citeauthoryear{{Tapping}}{{Tapping}}{1987}]{tapping1987}
{Tapping} K.~F.,  1987, \mn@doi [\jgr] {10.1029/JD092iD01p00829}, \href
  {http://adsabs.harvard.edu/abs/1987JGR....92..829T} {92, 829}

\bibitem[\protect\citeauthoryear{Tjorve \& Tjorve}{Tjorve \&
  Tjorve}{2017}]{tjorve2017}
Tjorve K. M.~C.,  Tjorve E.,  2017, \mn@doi [PLOS ONE]
  {10.1371/journal.pone.0178691}, 12, 1

\bibitem[\protect\citeauthoryear{{Usoskin}}{{Usoskin}}{2017}]{usoskin2017}
{Usoskin} I.~G.,  2017, \mn@doi [Living Reviews in Solar Physics]
  {10.1007/s41116-017-0006-9}, \href
  {http://adsabs.harvard.edu/abs/2017LRSP...14....3U} {14, 3}

\bibitem[\protect\citeauthoryear{{Woodard} \& {Noyes}}{{Woodard} \&
  {Noyes}}{1985}]{woodard1985}
{Woodard} M.~F.,  {Noyes} R.~W.,  1985, \mn@doi [\nat] {10.1038/318449a0},
  \href {http://adsabs.harvard.edu/abs/1985Natur.318..449W} {318, 449}

\makeatother
\end{thebibliography}



\bsp	
\label{lastpage}
\end{document}